\documentclass[aps,prb,reprint,groupedaddress]{revtex4-1}
\usepackage[dvipdfmx]{graphicx}
\usepackage{amsmath}
\usepackage{amssymb}
\usepackage{mathtools}
\usepackage{mathcomp}
\usepackage{float}
\usepackage{bm}
\usepackage{color}
\usepackage{braket}

\begin{document}

% Use the \preprint command to place your local institutional report
% number in the upper righthand corner of the title page in preprint mode.
% Multiple \preprint commands are allowed.
% Use the 'preprintnumbers' class option to override journal defaults
% to display numbers if necessary
%\preprint{}

%Title of paper
\title{
Chiral-phonon-induced current in helical crystals
}

% repeat the \author .. \affiliation  etc. as needed
% \email, \thanks, \homepage, \altaffiliation all apply to the current
% author. Explanatory text should go in the []'s, actual e-mail
% address or url should go in the {}'s for \email and \homepage.
% Please use the appropriate macro foreach each type of information

% \affiliation command applies to all authors since the last
% \affiliation command. The \affiliation command should follow the
% other information
% \affiliation can be followed by \email, \homepage, \thanks as well.
\author{Dapeng Yao$^{1}$ and Shuichi Murakami$^{1,2}$}
%\email[]{Your e-mail address}
%\homepage[]{Your web page}
%\thanks{}
%\altaffiliation{}
\affiliation{$^{1}$Department of Physics, Tokyo Institute of Technology, 2-12-1 Ookayama, Meguro-ku, Tokyo 152-8551, Japan \\
$^{2}$TIES, Tokyo Institute of Technology, Ookayama, Meguro-ku, Tokyo 152-8551, Japan}
%Collaboration name if desired (requires use of superscriptaddress
%option in \documentclass). \noaffiliation is required (may also be
%used with the \author command).
%\collaboration can be followed by \email, \homepage, \thanks as well.
%\collaboration{}
%\noaffiliation

\date{\today}

\begin{abstract}
In this study, we theoretically show that in a helical crystal, a current is induced by chiral phonons representing the microscopic local rotation of atoms. By treating the rotational motion as a perturbation, we calculate the time-dependent current by using the adiabatic Berry phase method. The time average of the current along the helical axis becomes finite in the metallic phase but it vanishes in the insulating phase. On the other hand, the current in the hexagonal plane changes with time, but its time average vanishes due to the threefold rotation space-time symmetry. We show that the time evolutions of the current follow the space-time symmetries of the helical systems. Moreover, we explain the reason for the vanishing of the time average of the current in the insulating phase from the aspect of the Chern number in the parameter space.

\end{abstract}

\maketitle

\section{Introduction}

Chirality is a fundamental property which reveals symmetry breaking of elementary particles. In condensed matter physics, electronic states around Weyl points become chiral, and it leads to many unconventional transport phenomena, such as chiral anomaly~\cite{Son,TaAs}, Klein tunneling~\cite{Klein} and unconventional Landau bands~\cite{Landau}. Recently, the induced current involving chirality, such as the circular photogalvanic effect in Weyl semimetals without inversion and mirror symmetries has been proposed~\cite{CPGEWeyl}, and the helicity-dependent photocurrent in topological insulators has been discovered~\cite{CPGETI}. This effect is a second-order chiral response which is generated by circular polarized light distinguished by a left-handed and a right-handed polarizations. It is similar to the chiral phonons at $\bm{K}$ and $\bm{K'}$ valleys in which the exchange of phonon angular momentum is excited through absorbing circularly polarized photon and emitting a chiral valley phonon~\cite{PMA15,CPexp}. 

A similar kind of chiral nature of eigenmodes in $\bm{k}$ space also appears in phonons. Studies of chiral phonons in the honeycomb lattice have revealed that phonons can have a microscopic local rotation of atoms in crystals, and this rotation is formulated as an angular momentum of phonons~\cite{PMA14,PMA22}. Various phenomena related to the phonon angular momentum have been investigated, such as the orbital magnetic moments of phonons~\cite{OMP1,OMP2,OMP3,OMP4,OMP5,OMP6}, magnon-phonon interconversion~\cite{MPItheo,MPIexp}, the phonon Hall effect~\cite{PHE1,PHE2,PHE3,PHE4,PHE5} and spin-phonon processes~\cite{SP1,SP2}. In particular, phonon-induced magnetization related to spin has been vigorously pursued for several years. One can show that in the presence of spin-orbit coupling, the microscopic local rotation can induce magnetization~\cite{Hamada,Geometric}. On the other hand, it has been reported that in a helical crystal structure, a magnetization along a helical axis can be induced by an electric current~\cite{Yoda1,Yoda2,Hara}. 
Combining the fact between the chiral-phonon-induced magnetization and the current-induced magnetization, a natural equation arises: is it possible that a magnetization and a current are coupled together when the chiral phonons meet the helical crystal structure with chirality? Because a magnetization and a current satisfy the similar symmetry properties~\cite{MJ,MJexp}, we expect that a rotational motions of atoms can induce a current in a helical system. 
In order to exhibit chiral-phonon-induced current, inversion symmetry should be broken. Among various systems without inversion symmetry, in this paper we choose helical crystals as an example, because helical systems, having right-handed and left-handed ones, are good for discussing various symmetry properties.

In this paper, we theoretically propose that in a helical crystal with honeycomb-lattice layers, a current along the helical axis is induced by the microscopic local rotation of atoms. We introduce a simple toy model which describes a helical system comprising layers of honeycomb lattices and suppose that the atoms are rotating with time due to a chiral phonon. The coupling between chiral phonons and electrons can be viewed as an adiabatic process, in which the electronic states related to Berry phase. There has been a substantial amount of study on chiral phonons, as well as effects arising from Berry phase treatment of chiral phonons in the last few years~\cite{Berry1,Berry2,Berry3,Berry4,Berry5}.
In the present paper, by treating the rotational motion as an adiabatic perturbation, we use the Berry phase method to calculate the time-dependent current. As a result, we obtain a finite chiral-phonon-induced current along the helical axis, and within the hexagonal plane its time average becomes zero in the metallic phases. On the other hand, in the insulating phases, the time average of the current along the helical axis vanishes and it can be explained naturally from the zero Chern number in the parameter space.

The remainder of the paper is organized as follows. In Sec.~\ref{secII}, we introduce a tight-binding model for a helical system with chiral phonons. Section~\ref{secIII} presents the calculation of the chiral-phonon-induced current. In Sec.~\ref{secIV}, we discuss dependence of the chiral-phonon-induced current on the onsite energy and the reason for vanishing of the current in insulators. Then we propose a realization of Thouless charge pumping. Section~\ref{secV} concludes this paper. Some details of our calculations are placed in Appendix.

\section{MODEL FOR A HELICAL SYSTEM WITH PHONON ANGULAR MOMENTUM}
\label{secII}
\subsection{Toy model with helical structure}
\begin{figure}[htb]
%\begin{figure*}[htb]
\begin{center}
\includegraphics[clip,width=8cm]{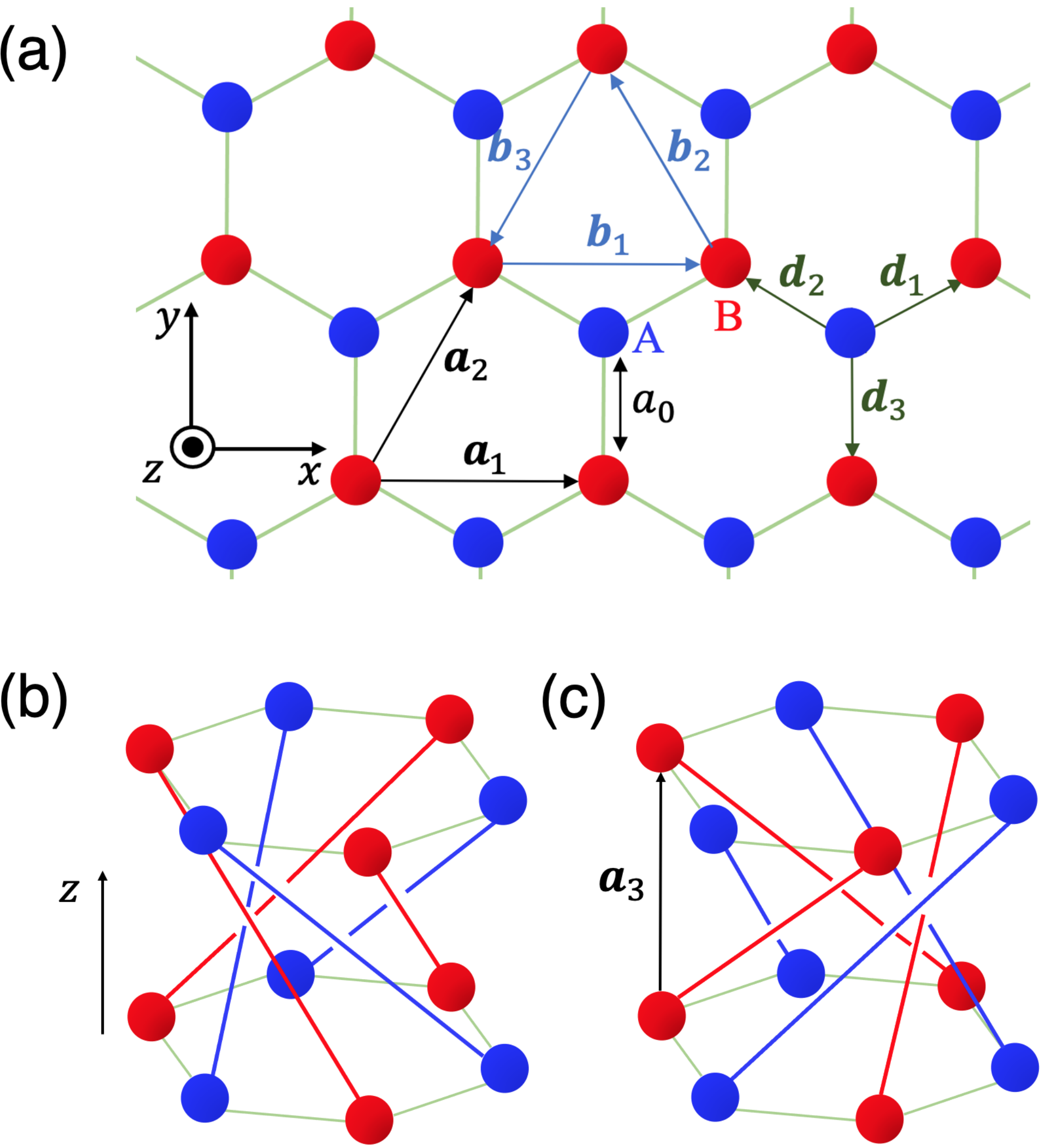}
\end{center}
\caption{
Helical structure with honeycomb-lattice layers. (a) One layer of the helical lattice, forming a honeycomb lattice. (b) Hopping pattern in the left-handed helical crystal. (c) Hopping pattern in the right-handed helical crystal. 
}
\label{Struct}
\end{figure}

First, we assume a helical crystal structure which is composed of an infinite stack of honeycomb lattice layers as shown in Fig.~\ref{Struct} with one $s$-like orbital per site. In Fig.~\ref{Struct}(a), the red and blue balls represent the A and B atoms at the two sublattices, respectively. The primitive translation vectors are chosen as $\bm{a}_1=a(1,0,0)$, $\bm{a}_2=a(1/2,\sqrt{3}/2,0)$, $\bm{a}_3=(0,0,c)$ with lattice constants $a$ in the $xy$ plane and $c$ along the $z$ direction. The vectors which connect the nearest-neighbor atoms are $\bm{d}_1=a_0(\sqrt{3}/2,1/2,0)$, $\bm{d}_2=a_0(-\sqrt{3}/2,1/2,0)$, $\bm{d}_3=a_0(0,-1,0)$, where $a_0=a/\sqrt{3}$ represents the length of the nearest-neighbor bond. For convenience, we label the vectors $\bm{b}_1=\bm{a}_1$, $\bm{b}_2=\bm{a}_2-\bm{a}_1$ and $\bm{b}_3=-\bm{a}_2$, which connect the next-nearest neighbor atoms. We introduce a three-dimensional tight-binding model with intra- and interlayer hoppings of electrons. Here, we consider a spinless tight-binding Hamiltonian for electrons given by
\begin{equation}
\label{HR}
\hat{H}_0=t_1\sum_{\braket{i,j}}\hat{c}_{i}^{\dagger}\hat{c}_{j}+\sum_{\mu=\rm{A,B}}\sum_{[i,j]}t_{\mu}\hat{c}_{\mu,i}^{\dagger}\hat{c}_{\mu,j}+\lambda_{\nu}\sum_{i}\xi_{i}\hat{c}_{i}^{\dagger}\hat{c}_{i},
\end{equation}
where $t_1$, $t_A$, $t_B$, $\lambda_{\nu}$ are real parameters and $\hat{c}_{i}~(\hat{c}_{i}^{\dagger})$ refers to an annihilation (a creation) operator at site $i$. The first term represents the nearest-neighbor hoppings within the honeycomb lattice layers. The second term represents the ``helical" hoppings between the neighboring honeycomb layers. In this term we put a subscript $\mu=(\rm{A,B})$ to distinguish two sublattices. We consider two models having different patterns of the helical hoppings, a right-handed and a left-handed ones. In the right-handed helical hoppings shown in Fig.~\ref{Struct}(b), the directions of interlayer hoppings between A sites are $\pm(-\bm{b}_i+\bm{a}_3)$ and those between B sites are $\pm(\bm{b}_i+\bm{a}_3)$. Similarly, in the model with the left-handed helical hoppings shown Fig.~\ref{Struct}(c), the directions of hoppings between A sites are $\pm(\bm{b}_i+\bm{a}_3)$ and those between B sites are $\pm(-\bm{b}_i+\bm{a}_3)$. Meanwhile, we suppose that the atoms A and B have the different helical hopping parameters $t_{\mu}, \mu=\rm{A,B}$. The existence of the helical hopping term breaks inversion and mirror symmetries which are held by the structure consisting of the honeycomb lattice layers without the helical hoppings. The third term is a staggered onsite energy $\xi_{i}\lambda_{\nu}$, where $\xi_i$ is $+1$ and $-1$ for A and B sublattices respectively. 

%---------------------------------------------------
%	Fig. 1: 
%---------------------------------------------------
%\end{figure*}
%---------------------------------------------------

Let $\mathcal{H}^{R}_{0}(\bm{k})$ and $\mathcal{H}^{L}_{0}(\bm{k})$ denote the $\bm{k}$-dependent Bloch Hamiltonians with the right-handed and left-handed hoppings, respectively. The Bloch Hamiltonians are represented as 
\begin{equation}
\label{BlochHam}
\mathcal{H}_0^{\alpha}(\bm{k})=d_0^{\alpha}(\bm{k})\sigma_0+d_x(\bm{k})\sigma_x+d_y(\bm{k})\sigma_y+d_z^{\alpha}(\bm{k})\sigma_z,
\end{equation}
where $\sigma_0$ is a $2\times2$ identity matrix, $\sigma_i$ are Pauli matrices, and $\alpha=R,L$ represents the two helical structures. Here the functions $d_i(\bm{k})~(i=x,y)$ and $d_j^{\alpha}(\bm{k})~(j=0,z)$ are 
\begin{align}
\label{d0}
d_0^{R,L}(\bm{k})=~&(t_A+t_B)\cos(k_zc)\sum_{i=1}^3\cos{(\bm{k}\cdot\bm{b}_i)} \nonumber \\
                      &\pm(t_A-t_B)\sin(k_zc)\sum_{i=1}^3\sin{(\bm{k}\cdot\bm{b}_i)}, \\ 
d_x(\bm{k})=~&t_1[1+C_1(\bm{k})+C_2(\bm{k})], \\
d_y(\bm{k})=~&t_1[S_1(\bm{k})+S_2(\bm{k})], \\
\label{dz}
d_z^{R,L}(\bm{k})=~&\lambda_{\nu}+(t_A-t_B)\cos(k_zc)\sum_{i=1}^3\cos{(\bm{k}\cdot\bm{b}_i)}  \nonumber \\   
                      &\pm(t_A+t_B)\sin(k_zc)\sum_{i=1}^3\sin{(\bm{k}\cdot\bm{b}_i)}, 
\end{align}
where
\begin{equation*}
C_1(\bm{k})=\cos{(\bm{k}\cdot\bm{a}_1)}, C_2(\bm{k})=\cos{(\bm{k}\cdot\bm{a}_2)}, 
\end{equation*}
\begin{equation}
S_1(\bm{k})=\sin{(\bm{k}\cdot\bm{a}_1)}, S_2(\bm{k})=\sin{(\bm{k}\cdot\bm{a}_2)}.
\end{equation}
More details about the connection between the tight-binding Hamiltonian Eq.~(\ref{HR}) and the Bloch Hamiltonian Eq.~(\ref{BlochHam}) are placed in Appendix. We notice that the helix is distinguished by the functions $d_z^{\alpha}(\bm{k})$ and $d_0^{\alpha}(\bm{k})$, in which $\alpha=R,L$ are corresponding to the right-handed and left-handed helical structures, respectively.

%---------------------------------------------------
%	Fig. 2: 
%---------------------------------------------------
\begin{figure}[htb]
\begin{center}
\includegraphics[clip,width=8.5cm]{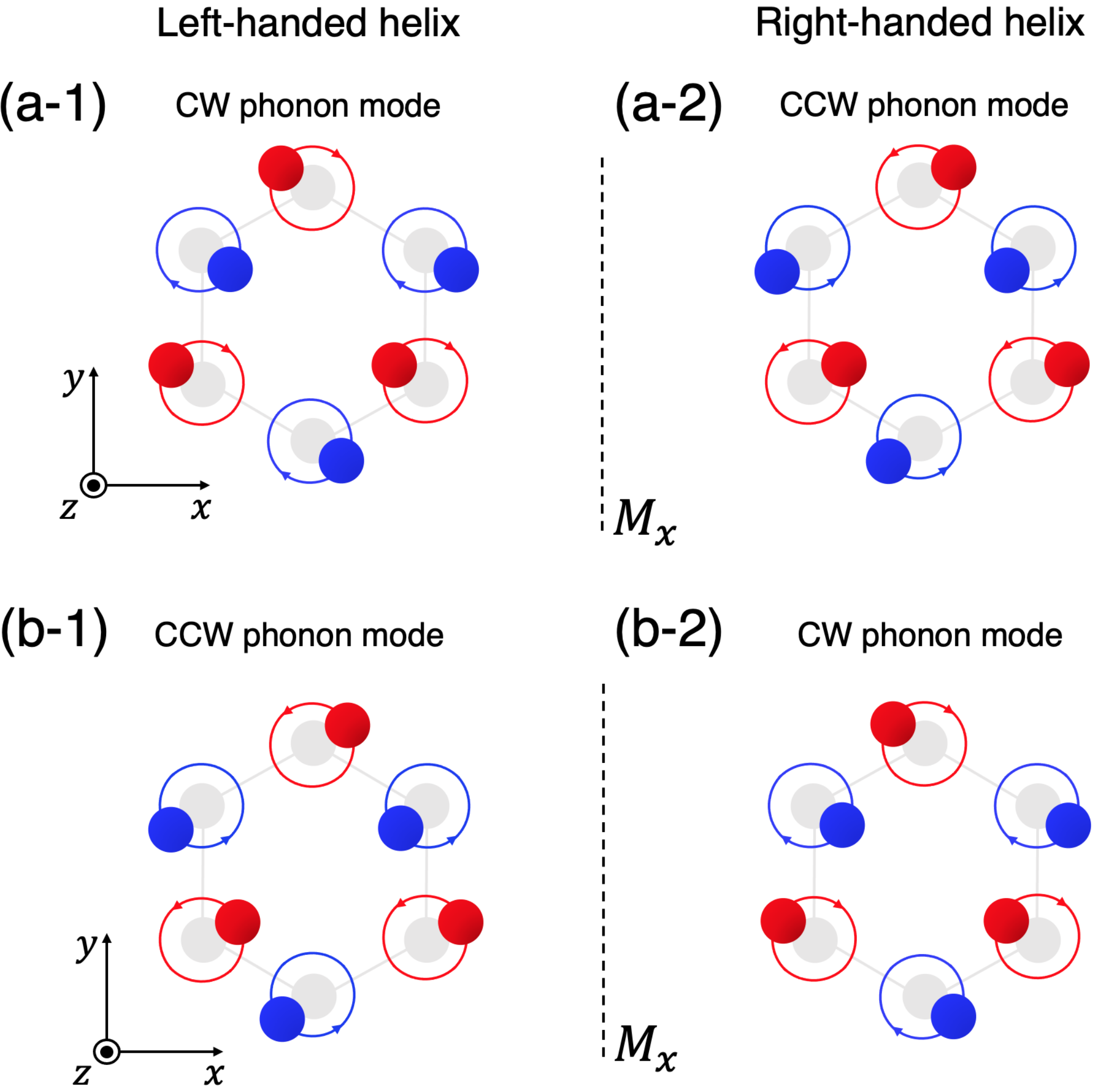}
\end{center}
\caption{
Two modes of microscopic rotation for the left-handed and right-handed helices related by the mirror reflection $M_x$ with respect to the $yz$ plane. (a-1) The left-handed helix with the CW (clockwise) phonon mode. (a-2) The right-handed helix with the CCW (counterclockwise) phonon mode. (b-1) The left-handed helix with the CCW phonon mode. (b-2) The right-handed helix with the CW phonon mode.}
\label{rotation}
\end{figure}
%---------------------------------------------------

\subsection{Microscopic local rotation as a perturbation}
Next, we introduce a perturbation term due to the microscopic local rotation of atoms as a special phonon mode. Here we consider chiral phonons at the $\bm{\Gamma}$ point of the Brillouin zone, which have been treated for the study of magnetization in recent years~\cite{OMP5,Hamada,GaCP1,GaCP2,GaCP3,GaCP4}.
At the $\bm{\Gamma}$ point, two optical phonon modes with displacement within the $xy$ plane are degenerate, and they are circularly polarized phonon modes corresponding to the microscopic local rotation of atoms. In this case of the phonon at the $\bm{\Gamma}$ point, the atoms in the same sublattices rotate with the same phase in the hexagonal plane, and the phase difference of the rotational motion between atoms A and B is $\pi$. In general, the rotational motion of atoms around the equilibrium position has two directions, the clockwise (CW) rotation and the counterclockwise (CCW) rotation. When the rotational motions within the hexagonal plane appear in the helical systems, we have four possibilities shown in Fig.~\ref{rotation}. Here we only consider the patterns in Figs.~\ref{rotation}(a-1) and~\ref{rotation}(a-2). The patterns in Figs.~\ref{rotation}(b-1) and ~\ref{rotation}(b-2) are obtained from Figs.~\ref{rotation}(a-1) and~\ref{rotation}(a-2) by the twofold rotation $C_{2y}$ with respect to the $y$ axis.

In Fig.~\ref{rotation}(a-1), we put the displacement vectors in the left-handed helix of atom A and atom B at time $t$ to be
\begin{equation}
\bm{r}_{A}^L=r_0^A(-\cos{\omega t},\sin{\omega t}), \bm{r}_{B}^L=-r_0^B(-\cos{\omega t},\sin{\omega t}),
\end{equation}
where $L$ labels the left-handed helix, and the angular velocity corresponding to the rotation of atoms is set to be $\omega$. The displacement vector from atom A to atom B is
\begin{equation}
\bm{r}^L=\bm{r}_B^L-\bm{r}_A^L=-r_{+}(-\cos{\omega t},\sin{\omega t}),
\end{equation}
with $r_{+}=r_0^A+r_0^B$. Then we assume that the lattice deformation $\bm{r}^{L}$ modifies the nearest-neighbor hopping parameter as $t_1\rightarrow t_1+\delta t^L_{a}$ \cite{Hamada,GraphHopping}, where $a=1,2,3$ represents the three directions of $\bm{d}_a$ corresponding to the nearest-neighbor hoppings. Then the first term in the Hamiltonian Eq.~(\ref{HR}) obtains an extra term depending on the time $t$ due to the microscopic local rotation of atoms. The extra term is defined as
\begin{equation}
\label{extra}
\hat{H}_{h}^{L}=\sum_{\braket{i,j}}\delta t_{ij}^{L}(t)\hat{c}_{i}^{\dagger}\hat{c}_{j},
\end{equation}
where $\delta t_{ij}^{L}$ is the modulation of the nearest-neighbor hopping parameter. It is natural to assume that the tight-binding hopping parameters for the $s$-like orbital depend only on the distance between atoms, and we set the modulation to be  proportional to the length change along the direction of the nearest-neighbor hopping, which is given by $\bm{r}^{L}(t)\cdot(\bm{d}_a/a_0)$. Thus, the modulation of the nearest-neighbor hopping parameter is written as
\begin{equation}
\delta t_{a}^{L}(t)=-\frac{t_1}{a_0^2}\bm{r}^{L}(t)\cdot\bm{d}_{a}.
\end{equation}
Therefore, the modulated Bloch Hamiltonian is given by
\begin{equation}
\mathcal{H}_h^{L}(\bm{k},t)=\mathcal{H}_c(\bm{k})\delta t\cos{\omega t}+\mathcal{H}_s(\bm{k})\delta t\sin{\omega t},
\end{equation}
where,
\begin{equation*}
\mathcal{H}_c(\bm{k})=\frac{\sqrt{3}}{2}\Bigl\{-[1-C_1(\bm{k})]\sigma_x+S_1(\bm{k})\sigma_y\Bigr\},
\end{equation*}
\begin{equation}
\mathcal{H}_s(\bm{k})=\frac{1}{2}\Bigl\{[1+C_1(\bm{k})-2C_2(\bm{k})]\sigma_x+[S_1(\bm{k})-2S_2(\bm{k})]\sigma_y\Bigr\},
\end{equation}
with $\delta t=r_{\pm}t_1/a_0$. 

Next, we define the displacement vector $\bm{r}^R$ in the right-handed helix with CCW phonons in Fig.~\ref{rotation}(a-2). We choose $\bm{r}^R=-r_{+}(\cos{\omega t},\sin{\omega t})$ so that it should be related to Fig.~\ref{rotation}(a-2) by the mirror symmetry $M_x$ with respect to the $yz$ plane. 

Finally, the total Bloch Hamiltonian considering a periodic microscopic local rotation of atoms for the left-handed helix with CW phonons depicted in Fig.~\ref{rotation}(a-1) is
\begin{equation}
\label{BH-L}
\mathcal{H}^{L}(\bm{k},t)=\mathcal{H}_0^{L}(\bm{k})+\mathcal{H}_c(\bm{k})\delta t\cos{\omega t}+\mathcal{H}_s(\bm{k})\delta t\sin{\omega t},
\end{equation}
and similarly, the total Bloch Hamiltonian for the right-handed helix with CCW phonons depicted in Fig.~\ref{rotation}(a-2) is written as
\begin{equation}
\label{BH-R}
\mathcal{H}^{R}(\bm{k},t)=\mathcal{H}_0^{R}(\bm{k})-\mathcal{H}_c(\bm{k})\delta t\cos{\omega t}+\mathcal{H}_s(\bm{k})\delta t\sin{\omega t}.
\end{equation}

\section{CHIRAL-PHONON-INDUCED CURRENT}
\label{secIII}
The motion of atoms in the chiral phonon mode will affect the electronic states. In the previous studies~\cite{OMP3,Berry3}, an adiabatic circular current produced by chiral phonons has been studied. In this section, we study an adiabatic linear current arising from Berry phase treatment of chiral phonons. We calculate expectation values of the current for the time-dependent periodic Hamiltonian by using the Berry phase method~\cite{Geometric,Hamada}.

\subsection{Berry phases in adiabatic process}
For our purpose, we review the general method of calculating an expectation value of an operator $\hat{X}$, given by
\begin{equation}
\hat{X}(t)=\partial_{\lambda}\hat{H}_{\lambda}(t),
\end{equation} 
where we introduce a parameter $\lambda$ into the Hamiltonian, following Refs.~\cite{Geometric,Hamada}. We assume that the Hamiltonian $\hat{H}_{\lambda}(t)$ has a periodic adiabatic dependence on time $t$, and let $\ket{\psi_{n,\lambda}(t)}$ be its instantaneous eigenstate of the band $n$ at time $t$ with an instantaneous eigenvalue $E_{n,\lambda}(t)$, where $n$ is the band index. Suppose that the change of Hamiltonian $\hat{H}_{\lambda}(t)$ is slow enough that the band index $n$ does not change at all with the change of the Hamiltonian. It means that the energy gap $\Delta_{nm}$ between the band $n$ and another band $m$ satisfies the adiabatic condition $\Delta_{nm}T/\hbar\gg1$ with the time period $T=2\pi/\omega$. The expectation value of the operator $\hat{X}(t)$ is defined as
\begin{equation}
\label{Xop}
X\equiv\frac{1}{T}\int_0^Tdt~tr[\hat{\rho}(t)\hat{X}(t)],
\end{equation}
where $\hat{\rho}(t)$ is the density matrix obeying the von Neumann equation $i\hbar\partial_t\hat{\rho}(t)=[\hat{H}_{\lambda}(t),\hat{\rho}(t)]$. In the absence of the time evolution, we set the density matrix to be identical to a projection to the $n$-th instantaneous eigenstate $\ket{\psi_{n,\lambda}}\bra{\psi_{n,\lambda}}$ only. It acquires contributions from other states satisfying $\hat{H}_{\lambda}(t)\ket{\psi_{m,\lambda}(t)}=E_{m,\lambda}(t)\ket{\psi_{m,\lambda}(t)}$ due to the time evolution. The matrix elements of the density matrix between the bands $n$ and $m$ are given by
\begin{equation}
\bra{\psi_{m,\lambda}(t)}\hat{\rho}(t)\ket{\psi_{n,\lambda}(t)}=i\hbar\frac{\bra{\psi_{m,\lambda}(t)}\partial_t\ket{\psi_{n,\lambda}(t)}}{E_{m,\lambda}(t)-E_{n,\lambda}(t)}.
\end{equation}
Next, the matrix elements of $\hat{X}(t)=\partial_{\lambda}\hat{H}_{\lambda}(t)$ between the bands $n$ and $m$ are written as
\begin{align}
&\bra{\psi_{n,\lambda}(t)}\hat{X}(t)\ket{\psi_{m,\lambda}(t)} \nonumber \\
 =&[E_{m,\lambda}(t)-E_{n,\lambda}(t)]\bra{\psi_{n,\lambda}(t)}\partial_{\lambda}\ket{\psi_{m,\lambda}(t)}
\end{align}
by using the Sternheimer equation
\begin{align}
&\partial_{\lambda}[\hat{H}_{\lambda}(t)-E_{n,\lambda}(t)]\ket{\psi_{n,\lambda}(t)}  \nonumber \\
&=[E_{n,\lambda}(t)-\hat{H}_{\lambda}(t)]\ket{\partial_{\lambda}\psi_{n,\lambda}(t)}.
\end{align}

By combining these equations and summing over the occupied states $n$, the expectation of the operator $\hat{X}(t)$ is written as
\begin{equation}
X(t)=\sum_n^{\rm{occ}}\Bigl(\partial_{\lambda}E_{n,\lambda}(t)+\mathcal{F}_{n,\lambda}(t)\Bigr),
\end{equation}
where
\begin{align}
\mathcal{F}_{n,\lambda}(t)\equiv i\hbar\partial_t\braket{\psi_{n,\lambda}(t)|\partial_{\lambda}\psi_{n,\lambda}(t)}- \nonumber \\
i\hbar\partial_{\lambda}\braket{\psi_{n,\lambda}(t)|\partial_t\psi_{n,\lambda}(t)}
\end{align}
is the Berry curvature corresponding to the band $n$ in the $(t,\lambda)$ space and $\sum_n^{\rm{occ}}$ represents a sum over the occupied states. Since in the time-periodic systems we can choose the instantaneous eigenstates to satisfy $\ket{\psi_{n,\lambda}(t)}=\ket{\psi_{n,\lambda}(t+T)}$, the general expression is written as
\begin{equation}
X(t)=\sum_n^{\rm{occ}}\bigr(X_n^{\rm{inst}}(t)+X_n^{\rm{geom}}(t)\bigl),
\end{equation}
with
\begin{align}
\label{inst}
X_n^{\rm{inst}}(t)\equiv&\partial_{\lambda}E_{n,\lambda}(t) \nonumber \\
=&\bra{\psi_n(t)}\hat{X}(t)\ket{\psi_n(t)}, 
\end{align}
\begin{equation}
\label{geom}
X_n^{\rm{geom}}(t)\equiv\hbar\sum_{m(\neq n)}\Bigl\{\frac{\hat{X}_{nm}(\lambda,t)A_{mn}(\lambda,t)}{E_{n,\lambda}(t)-E_{m,\lambda}(t)}+\rm{C.c}\Bigr\},
\end{equation}
where $\hat{X}_{nm}(\lambda,t)=\bra{\psi_n(t)}\partial_{\lambda}\hat{H}(t)\ket{\psi_m(t)}$ and $A_{mn}(\lambda,t)=\bra{\psi_m(t)}(-i\partial_t)\ket{\psi_n(t)}$ are the matrix elements of the operator $\hat{X}(t)$ and the Berry phase, repectively.

\subsection{Chiral-phonon-induced current as an adiabatic process}
Now, we apply this method to calculate an expectation value of the current operator, which is written as the derivative of Bloch Hamiltonian $\mathcal{H}(\bm{k})$ with respect to the wave vector $\bm{k}$:
\begin{equation}
\hat{\bm{J}}=\frac{-e}{\hbar}\partial_{\bm{k}}\mathcal{H}(\bm{k}),
\end{equation}
where $-e$ is the electron charge. Here we assume that the phonon frequency is much smaller than the excitation gap at each $\bm{k}$, and therefore the coupling between phonons and electrons is viewed as an adiabatic process. This assumption will be discussed and justified later in Sec.~\ref{secIV}~A. We use the Berry phase method to calculate the expectation value of the current operator $\bm{\hat{J}}$. From Eq.~(\ref{inst}) and Eq.~(\ref{geom}), the expectation value of the current operator at a fixed $\bm{k}_0$ point for a multiband model is written as
\begin{widetext}
\begin{eqnarray}
\hat{\bm{J}}(t)|_{\bm{k}_0} = \sum_{n}^{\rm{occ}}\bra{\psi_{n,\bm{k}}(t)}\hat{\bm{J}}_{\bm{k}}(t)\ket{\psi_{n,\bm{k}}(t)}|_{\bm{k}_0} +\hbar\sum_{n}^{\rm{occ}}\sum_{m(\neq n)}\Bigl\{\frac{\hat{\bm{J}}_{nm}(\bm{k},t)A_{mn}(\bm{k},t)}{E_{n,\bm{k}}(t)-E_{m,\bm{k}}(t)}\Big|_{\bm{k}_0}+\rm{C.c}\Bigr\},
\end{eqnarray}
\end{widetext}
where $\hat{\bm{J}}_{nm}(\bm{k},t)=\bra{\psi_{n,\bm{k}}(t)}\hat{\bm{J}}_{\bm{k}}(t)\ket{\psi_{m,\bm{k}}(t)}$ and $A_{mn}(\bm{k},t)=\bra{\psi_{m,\bm{k}}(t)}(-i\partial_t)\ket{\psi_{n,\bm{k}}(t)}$ are the matrix elements of the current operator and the Berry connection, respectively. Considering the contributions from all the $\bm{k}$-points, the total current is expressed as a summation over the entire first Brillouin zone. Therefore, the total expectation value of the current operator is
\begin{equation}
\hat{\bm{J}}(\tau)=\sum_{n}^{\rm{occ}}[\bm{j}_n^{\rm{inst}}(\tau)+\bm{j}_n^{\rm{geom}}(\tau)],
\end{equation}
with
\begin{equation}
\label{Jinst}
\bm{j}_n^{\rm{inst}}(\tau)=\frac{-e}{V}\sum_{\bm{k}}\hat{\bm{v}}_{nn}(\bm{k},\tau),
\end{equation}
\begin{equation}
\label{Jgeom}
\bm{j}_n^{\rm{geom}}(\tau)=\frac{-e\omega}{V}\sum_{\bm{k}}\sum_{m(\neq n)}\Bigl\{\frac{\hat{\bm{v}}_{nm}(\bm{k},\tau)A_{mn}(\bm{k},\tau)}{E_{n,\bm{k}}(\tau)-E_{m,\bm{k}}(\tau)}+\rm{C.c}\Bigr\},
\end{equation}
where $V$ is the total volume of the crystal. We introduce a dimensionless quantity $\tau=\omega t$ to replace time $t$, where $\omega$ represents the phonon frequency. Here we define the matrix elements of the velocity $\hat{\bm{v}}_{nm}(\bm{k},\tau)=\bra{\psi_{n,\bm{k}}(\tau)}\partial_{\bm{k}}\mathcal{H}(\bm{k},\tau)\ket{\psi_{m,\bm{k}}(\tau)}$, and the Berry connection is rewritten as $A_{mn}(\bm{k},\tau)=\bra{\psi_{m,\bm{k}}(\tau)}(-i\partial_\tau)\ket{\psi_{n,\bm{k}}(\tau)}$. The expectation value of the current is divided into two parts; the first part $\bm{j}_n^{\rm{inst}}(\tau)$ is called instantaneous current for the band $n$ at the rescaled time $\tau$ and the second part $\bm{j}_n^{\rm{geom}}(\tau)$ is dubbed geometric current, which is associated with Berry connection in the adiabatic process. 

%---------------------------------------------------
%	Fig. 3: 
%---------------------------------------------------
\begin{figure}[htb]
\begin{center}
\includegraphics[clip,width=8.5cm]{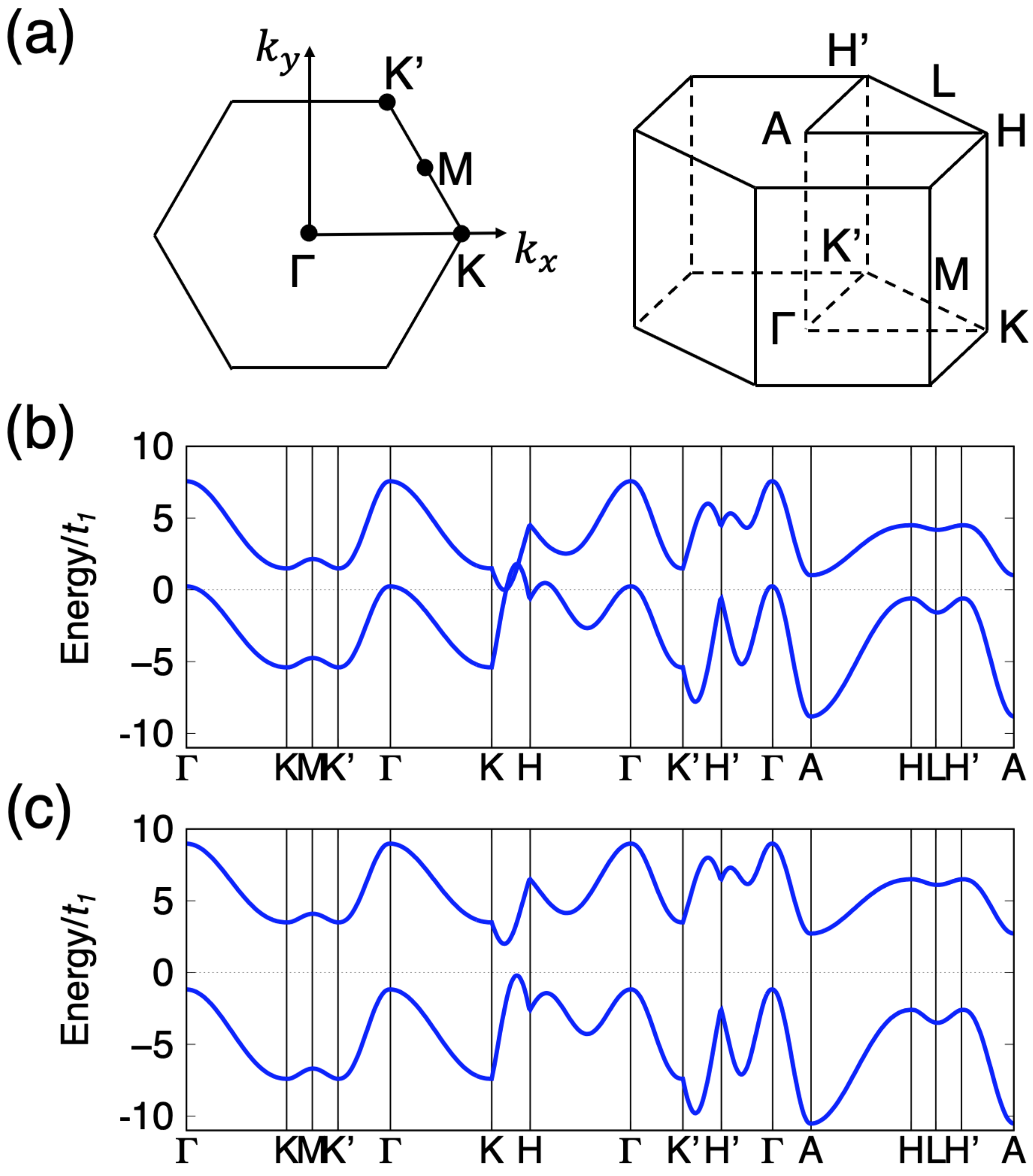}
\end{center}
\caption{
Band structure for the helical model of a stacked honeycomb lattice without phonons. The lattice constants are $a=1$, $c=2$ and the hopping parameters are set to be $t_A=0.5t_1$, $t_B=0.8t_1$. (a) The first Brillouin zone of the model. (b) The metallic band dispersion with the onsite energy value $\lambda_{\nu}=3t_1$. (c) The insulating band dispersion with the onsite energy value $\lambda_{\nu}=5t_1$.}
\label{bandplt}
\end{figure}
%---------------------------------------------------

\subsection{Symmetry analysis}
We consider symmetries in our model, and study symmetry properties of the current. The time-dependent Bloch Hamiltonian Eq.~(\ref{BH-L}) for the left-handed helix with CW phonons and Eq.~(\ref{BH-R}) for the right-handed helix with CCW phonons are rewritten as
\begin{equation}
\mathcal{H}^{L}(\bm{k},\tau)=\mathcal{H}_0^{L}(\bm{k})+\mathcal{H}_c(\bm{k})\delta t\cos{\tau}+\mathcal{H}_s(\bm{k})\delta t\sin{\tau},
\end{equation}
\begin{equation}
\mathcal{H}^{R}(\bm{k},\tau)=\mathcal{H}_0^{R}(\bm{k})-\mathcal{H}_c(\bm{k})\delta t\cos{\tau}+\mathcal{H}_s(\bm{k})\delta t\sin{\tau},
\end{equation}
with the rescaled time $\tau$. In the absence of phonons, only the spatial symmetries need to be considered. The inversion symmetry is broken by the staggered onsite energy in our model. As we mentioned before, the right-handed and left-handed structures are connected with each other by the mirror symmetry $M_x$ with respect to the $yz$ plane. On the other hand, each helical structure holds the threefold rotation symmetry $C_{3z}$ with respect to the $z$ axis because the layers are composed of honeycomb lattice within the $xy$ plane. In the presence of phonons, the lattice structure is deformed slightly and periodically with time. In this case, the spatial symmetries are no longer enough, and the extra space-time symmetries should be taken into account. 

The instantaneous Hamiltonian $\mathcal{H}^{\alpha}(\bm{k},\tau)~(\alpha=R,L)$ perserves the time-reversal symmetry, where the time-reversal operator for the spinless system is expressed as $\Theta=K$ with $K$ being the complex-conjugate operator. Namely, the time-dependent Bloch Hamiltonian satisfies
\begin{equation}
\Theta\mathcal{H}^{\alpha}(\bm{k},\tau)\Theta^{-1}=\mathcal{H}^{\alpha}(-\bm{k},\tau),~\alpha=R,L
\end{equation}
without changing the sign of the rescaled time $\tau$ because of $\Theta\mathcal{H}_{c(s)}(\bm{k})\Theta^{-1}=\mathcal{H}_{c(s)}(-\bm{k})$. Thus, the instantaneous eigenstates $\ket{\psi^{\alpha}_{n,\bm{k}}(\tau)}$ and the instantaneous eigenenergy $E^{\alpha}_{n,\bm{k}}(\tau)$ satisfy
\begin{equation}
\Theta\ket{\psi^{\alpha}_{n,\bm{k}}(\tau)}=\ket{\psi^{\alpha}_{n,-\bm{k}}(\tau)},~~E^{\alpha}_{n,\bm{k}}(\tau)=E^{\alpha}_{n,-\bm{k}}(\tau),
\end{equation}
which lead to the symmetry properties of the instantaneous matrix elements for the velocity $\hat{\bm{v}}^{\alpha}_{nm}(\bm{k},\tau)$ and the Berry connection $A^{\alpha}_{mn}(\bm{k},\tau)$ as
\begin{align}
\hat{\bm{v}}^{\alpha}_{nm}(\bm{k},\tau)&=\bra{\psi^{\alpha}_{n,\bm{k}}(\tau)}\partial_{\bm{k}}\mathcal{H}^{\alpha}(\bm{k},\tau)\ket{\psi^{\alpha}_{m,\bm{k}}(\tau)} \nonumber \\
                         &=\bra{\psi^{\alpha}_{n,-\bm{k}}(\tau)}(-\partial_{-\bm{k}}\mathcal{H}^{\alpha}(-\bm{k},\tau))\ket{\psi^{\alpha}_{m,-\bm{k}}(\tau)} \nonumber \\
                         &=-\hat{\bm{v}}^{\alpha}_{nm}(-\bm{k},\tau),
\end{align}
and
\begin{align}
A^{\alpha}_{mn}(\bm{k},\tau)&=\bra{\psi^{\alpha}_{m,\bm{k}}(\tau)}(-i\partial_{\tau})\ket{\psi^{\alpha}_{n,\bm{k}}(\tau)} \nonumber \\
                         &=\bra{\psi^{\alpha}_{m,-\bm{k}}(\tau)}(i\partial_{\tau})\ket{\psi^{\alpha}_{n,-\bm{k}}(\tau)} \nonumber \\
                         &=-A^{\alpha}_{mn}(-\bm{k},\tau).
\end{align}
From these relations, we directly notice that the instantaneous current $\bm{j}_n^{\rm{inst},\alpha}(\tau)$ in Eq.~(\ref{Jinst}) becomes zero since the terms at $\bm{k}$ and $-\bm{k}$ cancel in the summation over the first Brillouin zone. Hence, only the geometric current is non-trivial and it contributes to the time-averaged current induced by chiral phonons.

For the geometric current $\bm{j}_n^{\rm{geom}}(\tau)$ in Eq.~(\ref{Jgeom}), we note the following two space-time symmetries. Firstly, mirror reflection gives the relationship of the geometric current between the left-handed helix with CW phonons (Fig.~\ref{rotation}(a-1)) and the right-handed helix with CCW phonons (Fig.~\ref{rotation}(a-2)). At any rescaled time $\tau$, the mirror reflection transfers the position $\bm{r}^R(\tau)$ of atoms in the right-handed helix into the position $\bm{r}^L(\tau)$ in the left-handed helix. Secondly, the system preserves the space-time symmetry, which is represented by the threefold rotation operation followed by $\mp T/3$ time translation. The time-dependent Bloch Hamiltonian under these space-time symmetries imposes the following conditions. Firstly, the mirror reflection connects the right-handed and left-handed helices
\begin{equation}
U_1\mathcal{H}^R(k_x,k_y,k_z,\tau)U_1^{-1}=\mathcal{H}^L(-k_x,k_y,k_z,\tau),
\end{equation}
where
\begin{equation}
U_1=
\begin{pmatrix}
1 & 0 \\
0 & e^{-i\bm{k}\cdot\bm{a}_1} \\
\end{pmatrix}.
\end{equation}
Secondly, the threefold rotation operation gives
\begin{align}
U_2\mathcal{H}^{L}(\bm{k},\tau)U_2^{-1}=\mathcal{H}^{L}(C_{3z}\bm{k},\tau-\frac{2\pi}{3}), \\
U_2\mathcal{H}^{R}(\bm{k},\tau)U_2^{-1}=\mathcal{H}^{R}(C_{3z}\bm{k},\tau+\frac{2\pi}{3}),
\end{align}
where 
\begin{equation}
U_2=
\begin{pmatrix}
e^{i\bm{k}\cdot\bm{a}_2} & 0 \\
0 & 1 \\
\end{pmatrix},
\end{equation}
and $C_{3z}\bm{k}=(-k_x/2-\sqrt{3}k_y/2,\sqrt{3}k_x/2-k_y/2,k_z)$. 

Under these two space-time symmetries, the relations between the right-handed and left-handed helices for the instantaneous matrix elements of the velocity $\hat{\bm{v}}_{nm}$ and the Berry connection $A_{mn}$ are obtained. Firstly, the mirror reflection gives the instantaneous matrix elements of the velocity
\begin{align}
&v_{nm,x}^R(k_x,k_y,k_z,\tau)=-v_{nm,x}^L(-k_x,k_y,k_z,\tau), \\
&v_{nm,y}^R(k_x,k_y,k_z,\tau)=v_{nm,y}^L(-k_x,k_y,k_z,\tau), \\
&v_{nm,z}^R(k_x,k_y,k_z,\tau)=v_{nm,z}^L(-k_x,k_y,k_z,\tau), 
\end{align}
and the instantaneous Berry connection
\begin{equation}
A_{mn}^R(k_x,k_y,k_z,\tau)=A_{mn}^L(-k_x,k_y,k_z,\tau).
\end{equation}
Secondly, the rotation operation gives the instantaneous matrix elements of the velocity
\begin{equation}
\hat{\bm{v}}^{\alpha}_{nm}(\bm{k},\tau)=C_{3z}^{-1}\hat{\bm{v}}^{\alpha}_{nm}(C_{3z}\bm{k},\tau\mp\frac{2\pi}{3}),
\end{equation}
and the instantaneous Berry connection
\begin{equation}
A^{\alpha}_{mn}(\bm{k},\tau)=A^{\alpha}_{mn}(C_{3z}\bm{k},\tau\mp\frac{2\pi}{3}),
\end{equation}
with the minus (plus) sign for $\alpha=L~(\alpha=R)$. Therefore, the relations of the time-dependent geometric current between the two helical structures connected by the space-time mirror reflection satisfies
\begin{subequations}
\begin{equation}
\label{jx}
j_{n,x}^{\rm{geom},R}(\tau)=-j_{n,x}^{\rm{geom},L}(\tau),
\end{equation}
\begin{equation}
\label{jyz}
j_{n,y(z)}^{\rm{geom},R}(\tau)=j_{n,y(z)}^{\rm{geom},L}(\tau),
\end{equation}
\end{subequations}
and for each helical structure, the geometric current satisfies
\begin{equation}
\label{jC3}
\bm{j}_{n}^{\rm{geom,\alpha}}(\tau)=C_{3z}^{-1}\bm{j}_{n}^{\rm{geom,\alpha}}(\tau\mp\frac{2\pi}{3}).
\end{equation}

%---------------------------------------------------
%	Fig. 4: 
%---------------------------------------------------
\begin{figure}[htb]
\begin{center}
\includegraphics[clip,width=8.5cm]{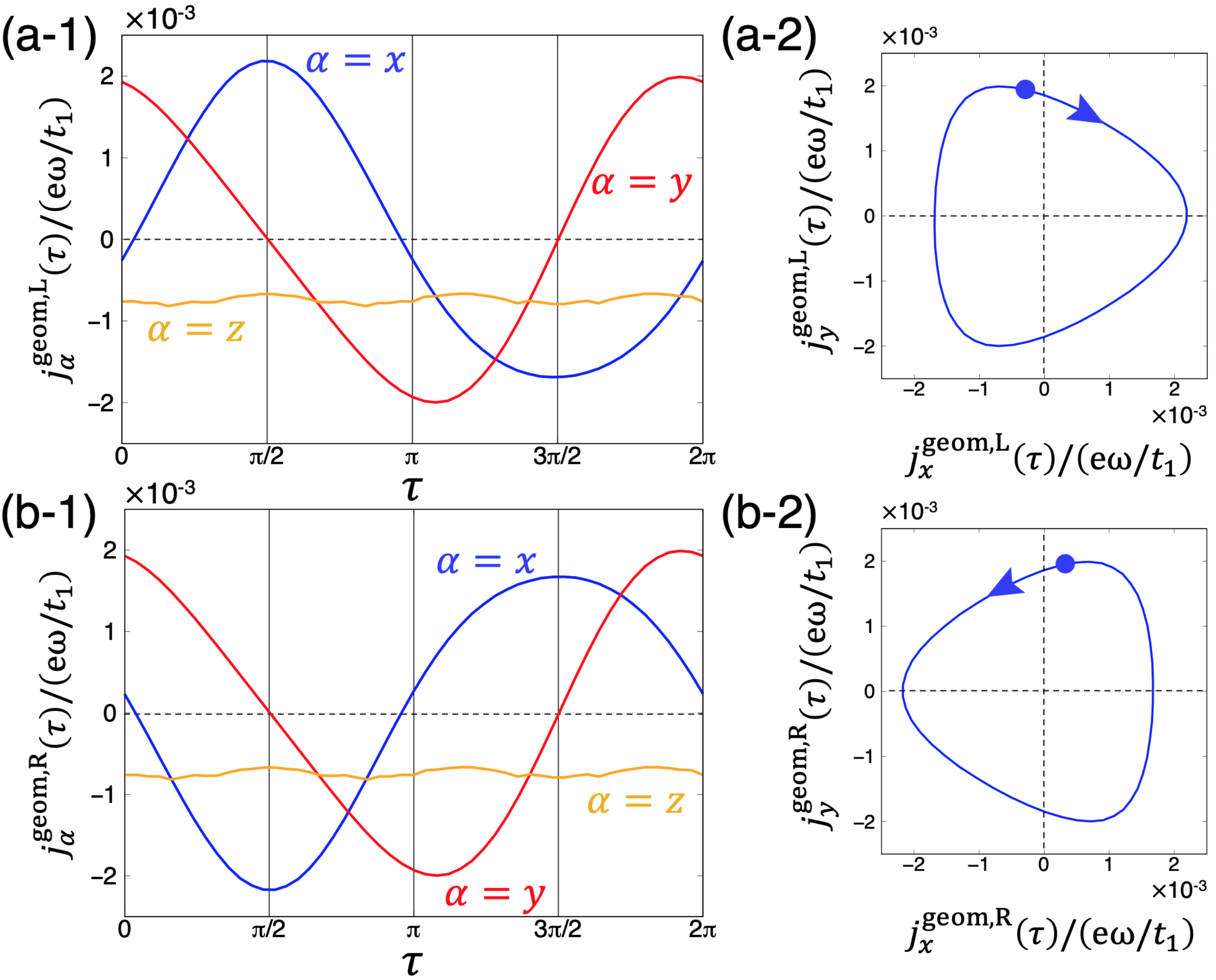}
\end{center}
\caption{
Results of the calculation of the geometric current for the metallic phases with $t_A=0.5$, $t_B=0.8$, $\delta t=0.1t_1$, and $\lambda_{\nu}=3t_1$. For the left-handed helix with CW phonons, (a-1) is the geometric current in $x,y,z$ directions and (a-2) represents the evolution of the geometric current for one period in the $xy$ plane. For the right-handed helix with CCW phonons, (b-1) is the geometric current in $x,y,z$ directions and (b-2) represents the evolution of the geometric current for one period in the $xy$ plane.}
\label{metal_current}
\end{figure}
%---------------------------------------------------

\begin{figure}[htb]
\begin{center}
\includegraphics[clip,width=8.5cm]{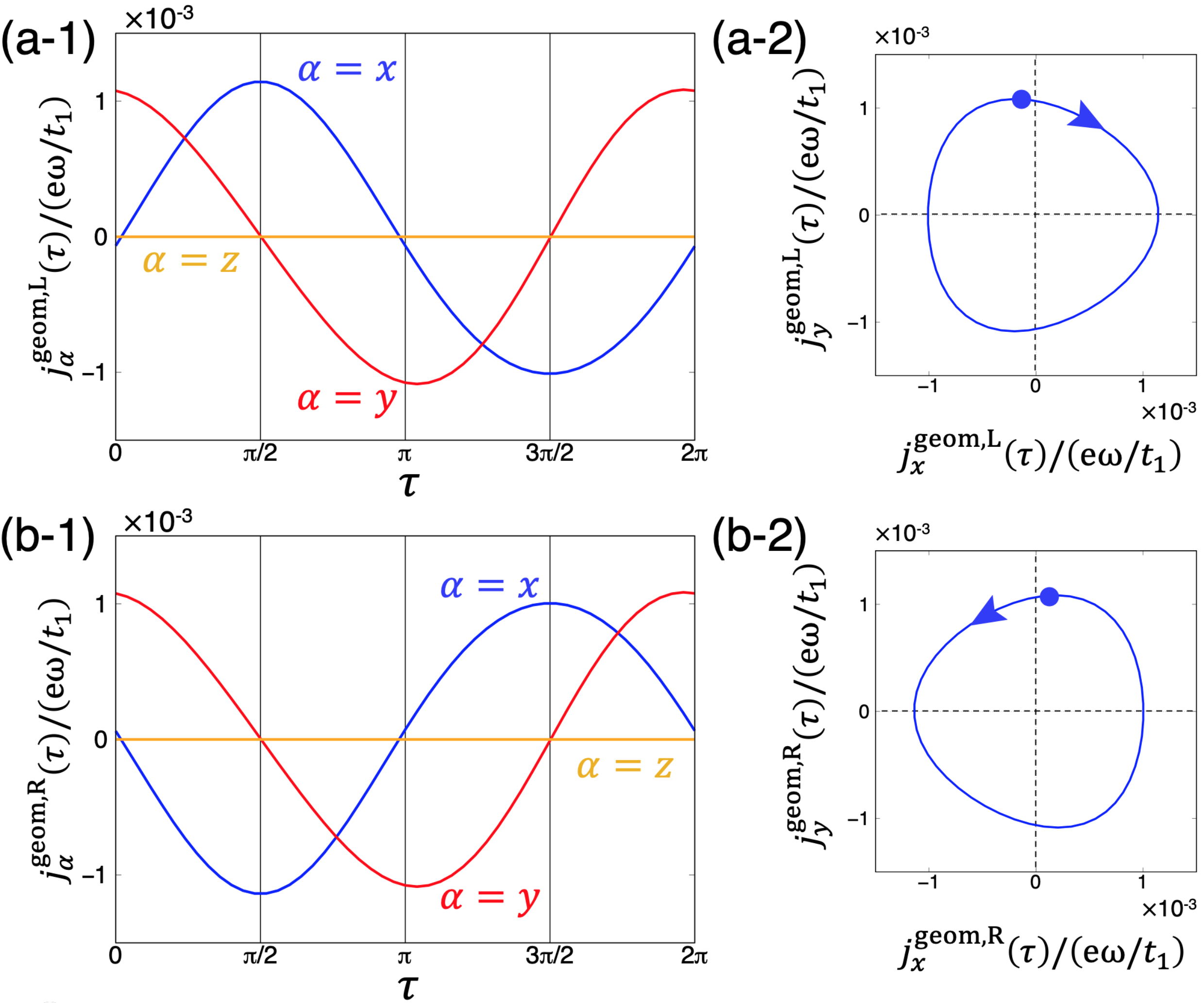}
\end{center}
\caption{
Results of the calculation of the geometric current for insulating phases with $t_A=0.5$, $t_B=0.8$, $\delta t=0.1t_1$ and $\lambda_{\nu}=5t_1$. For the left-handed helix with CW phonons, (a-1) is the geometric current in $x,y,z$ directions and (a-2) represents the evolution of the geometric current for one period in the $xy$ plane. For the right-handed helix with CCW phonons, (b-1) is the geometric current in $x,y,z$ directions and (b-2) represents the evolution of the geometric current for one period in the $xy$ plane.}
\label{insulator_current}
\end{figure}

We numerically calculate the geometric current $\bm{j}^{\rm{geom}}(\tau)$ for our two-band spinless model. The electronic energy band structure along the high symmetry points (see Fig.~\ref{bandplt}(a)) without phonons is shown in Figs.~\ref{bandplt}(b) and~\ref{bandplt}(c), whose onsite energy is set to be $\lambda_{\nu}=3t_1$ and $\lambda_{\nu}=5t_1$, representing metallic and insulating phases, respectively. 
Meanwhile, we set the modulated hopping term arising from chiral phonons to be $\delta t=0.1t_1$, which is smaller than other parameters and can viewed as a perturbation term to calculate the induced currents. It means that the displacement is around $10\%$ of the lattice constant.
We plot the geometric current for the metallic phases with $\lambda_{\nu}=3t_1$ in Fig.~\ref{metal_current} and that for the insulating phases with $\lambda_{\nu}=5t_1$ in Fig.~\ref{insulator_current}. We compare the time-evolution of the geometric current between two opposite helical structures. We see that the symmetry properties in Eqs.~(\ref{jx}),~(\ref{jyz}) and~(\ref{jC3}), are indeed satisfied. On the other hand, in the $xy$ plane, the geometric currents rotate and its trajectory forms a triangle-like cycle as shown in Figs.~\ref{metal_current}(a-2),~\ref{metal_current}(b-2) and Figs.~\ref{insulator_current}(a-2),~\ref{insulator_current}(b-2) for the metallic and insulating phases, respectively. The time average of the geometric currents in the $xy$ plane over one period vanishes due to the space-time threefold rotation symmetry. Here we notice that the time-averaged $z$ component of the geometric current is finite for the metallic phases but it vanishes for the insulating phases.

%---------------------------------------------------
%	Fig. 5: 
%---------------------------------------------------
%---------------------------------------------------

\section{DISCUSSION}
\label{secIV}
In this section, we discuss two aspects of the chiral-phonon-induced current. The first one is the dependence of the geometric current on the onsite energy $\lambda_{\nu}$. The second aspect is to explain why the time-averaged $z$ component of the geometric current vanishes in the insulating phases. More precisely, we show that this time-averaged current in insulators with a periodic change of parameters is quantized, but in systems with chiral phonons the quantized value is zero. 
To illustrate this quantization, we propose a toy-model realization of the charge pumping apart from systems with chiral phonons.
\subsection{Dependence of the chiral-phonon-induced current on the onsite energy $\lambda_{\nu}$}
%---------------------------------------------------
%	Fig. 6: 
%---------------------------------------------------
%\begin{figure}[htb]
\begin{figure*}[htb]
\begin{center}
\includegraphics[clip,width=18cm]{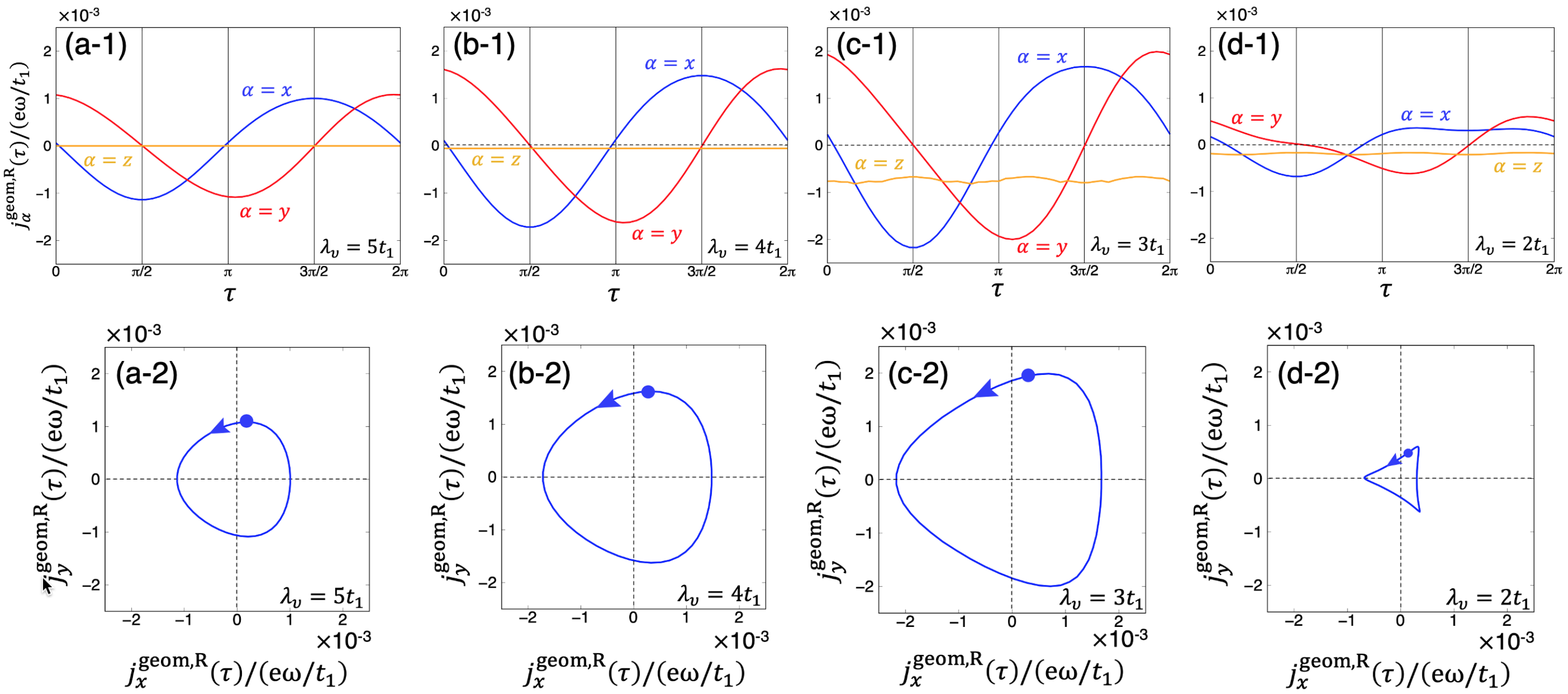}
\end{center}
\caption{
Dependence of the geometric current on the onsite energy $\lambda_{\nu}$ with the parameter values $t_A=0.5$, $t_B=0.8$ and $\delta t=0.1t_1$ for the right-handed helix with CCW phonons. The onsite energies are set to be (a-1), (a-2): $\lambda_{\nu}=5t_1$, (b-1), (b-2): $\lambda_{\nu}=4t_1$, (c-1), (c-2): $\lambda_{\nu}=3t_1$, and (d-1), (d-2): $\lambda_{\nu}=2t_1$. (a-1)-(d-1) are the results of the geometric current induced by the chiral phonon. (a-2)-(d-2) are the trajectories of the current vector $\bm{j}^{\rm{geom},R}$ within the $xy$ plane. The arrows show the direction of time evolution and these trajectories form the triangle-like cycles.
}
\label{cur_on}
%\end{figure}
\end{figure*}
%---------------------------------------------------

Here we focus on the right-handed helix with CCW phonons, and the geometric current for the onsite energies $\lambda_{\nu}=5t_1,~4t_1,~3t_1,~2t_1$ are shown in Figs.~\ref{cur_on}(a-1)-\ref{cur_on}(d-1). In Fig.~\ref{cur_on}(a-1)($\lambda_{\nu}=5t_1$), the $z$ component of the geometric current vanishes because it is an insulator. As the onsite energy decreases, the two energy bands gradually become close to each other in the Brillouin zone and the system becomes metallic. In the metallic phases, the $z$ component of the geometric current becomes finite. On the other hand, the $x$ and $y$ components rotate in the $xy$ plane with a triangle-like trajectory due to the threefold rotation symmetry. The shape of the trajectory formed by the $x$ and $y$ components changes with the change of the onsite energy while preserving the threefold rotation symmetry under the time evolution.

%---------------------------------------------------
%	Fig. 7: 
%---------------------------------------------------
%\begin{figure}[htb]
\begin{figure*}[htb]
\begin{center}
\includegraphics[clip,width=18cm]{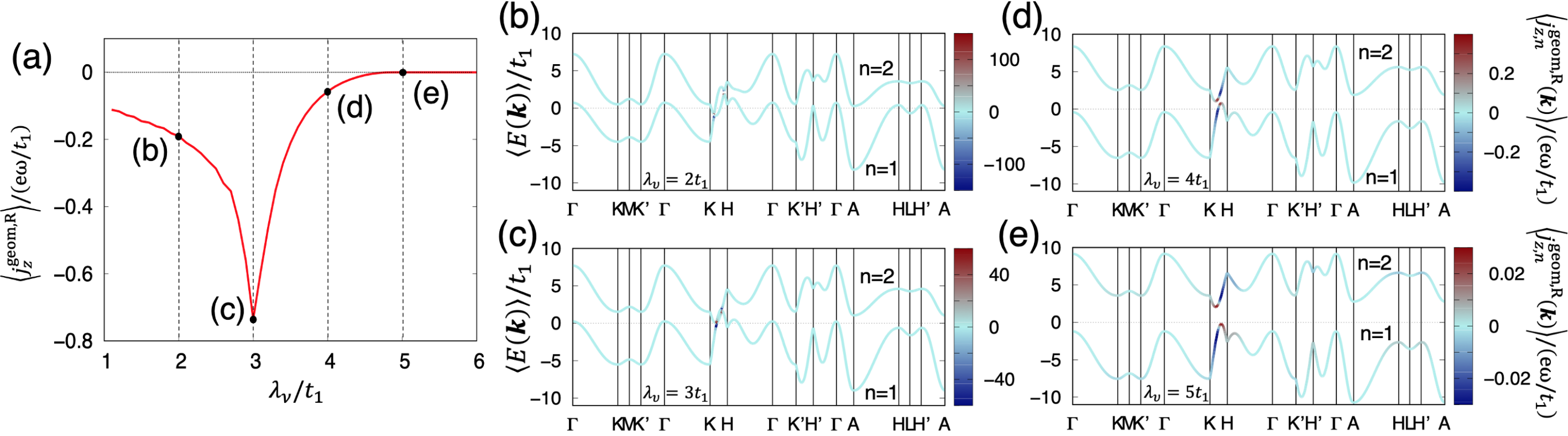}
\end{center}
\caption{
Contribution to the $z$ component of the geometric current from the respective time-averaged energy band with the parameter values $t_A=0.5$, $t_B=0.8$ and $\delta t=0.1t_1$ for the right-handed helix with CCW phonons. (a) Dependence on the onsite energy of the $z$ component of time-averaged geometric current. (b)-(e) Time-averaged energy bands with the color map showing the contribution to the time-averaged geometric current. The onsite energies are set to be (b) $\lambda_{\nu}=2t_1$, (c) $\lambda_{\nu}=3t_1$, (d) $\lambda_{\nu}=4t_1$, and (e) $\lambda_{\nu}=5t_1$.
}
\label{onsite}
%\end{figure}
\end{figure*}
%---------------------------------------------------

Next, we pay more attention to the $z$ component of the geometric current and discuss how it depends on the band structure. In our model, the time-dependent perturbation cause the dynamical change of the energy band. The time-dependent geometric current can be rewritten as
\begin{align}
\label{discussJ}
&\sum_{n}^{\rm{occ}}\bm{j}_n^{\rm{geom}}(\tau) \nonumber \\
&=-e\omega\sum_{\bm{k}}\sum_{n}^{\rm{occ}}\sum_{m(\neq n)}\Bigl\{\frac{\hat{\bm{v}}_{nm}(\bm{k},\tau)A_{mn}(\bm{k},\tau)}{E_{n,\bm{k}}(\tau)-E_{m,\bm{k}}(\tau)}+\rm{C.c}\Bigr\} \nonumber \\
&=-e\omega\sum_{\bm{k}}\sum_{n}^{\rm{occ}}\sum_{m}^{\rm{unocc}}\Bigl\{\frac{\hat{\bm{v}}_{nm}(\bm{k},\tau)A_{mn}(\bm{k},\tau)}{E_{n,\bm{k}}(\tau)-E_{m,\bm{k}}(\tau)}+\rm{C.c}\Bigr\},
\end{align}
where $\sum_n^{\rm{unocc}}$ represents a sum over the unoccupied states $m$. Here, the summation over $m$ is replaced from that with $m\neq n$ to that over the unoccupied states. We can see that the geometric current is proportional to the inverse of the energy difference between the $n$-th occupied band and the $m$-th unoccupied band $[E_{n,\bm{k}}(\tau)-E_{m,\bm{k}}(\tau)]^{-1}$, and the magnitude of the geometric current becomes larger when the gap is small, as shown in Fig.~\ref{onsite}(a). Within the range of $\lambda_{\nu}>3t_1$, the band gap gradually becomes larger for a larger value of $\lambda_{\nu}$ so that the $z$ component of the geometric current decreases with the onsite energy. The magnitude of the geometric current, however is not only dependent on the energy difference but also the product of the matrix element for the velocity and the Berry connection in the numerator of Eq.~(\ref{discussJ}). Within the range of $\lambda_{\nu}<3t_1$, the $z$ component of the geometric current increases with an increase of the onsite energy. For these onsite energies $\lambda_{\nu}=2t_1,~3t_1,~4t_1,~5t_1$, we show the contribution to the $z$ component of the geometric current from the energy band in Figs.~\ref{onsite}(b)-\ref{onsite}(e). Here we show the time average of the energy band under the time-dependent perturbation. Namely, we plot the time-averaged energy eigenvalue $\langle E_n(\bm{k})\rangle$ along the high-symmetry lines, where
\begin{equation}
\langle E_n(\bm{k})\rangle=\frac{1}{2\pi}\int_{0}^{2\pi}d\tau\bra{\psi_{n,\bm{k}}(\tau)}\mathcal{H}(\bm{k},\tau)\ket{\psi_{n,\bm{k}}(\tau)},
\end{equation}
and the time-averaged geometric current for the $z$ component 
\begin{equation}
\label{aveJ}
\begin{split}
&\langle j_{z,n}^{\rm{geom}}(\bm{k})\rangle \\
&=\frac{-e\omega}{2\pi}\int_{0}^{2\pi}d\tau\sum_{m(\neq n)}\Bigl\{\frac{\hat{v}_{z,nm}(\bm{k},\tau)A_{mn}(\bm{k},\tau)}{E_{n,\bm{k}}(\tau)-E_{m,\bm{k}}(\tau)}+\rm{C.c}\Bigr\}. 
\end{split}
\end{equation}
The results show that in the metallic phases, the $z$ component of the geometric current comes from the band degeneracy along the high-symmetry line $\overline{KH}$, in which the denominator in the Eq.~(\ref{aveJ}) vanishes and the magnitude $\langle\bm{j}_{z,n}^{\rm{geom}}(\tau)\rangle$ becomes larger as shown in Figs.~\ref{onsite}(b),~\ref{onsite}(c). Once the onsite energy is too large to let the band touch as shown in Figs.~\ref{onsite}(d),~\ref{onsite}(e), the magnitude $\langle\bm{j}_{z,n}^{\rm{geom}}(\tau)\rangle$ becomes smaller until the $z$ component of the geometric current eventually vanishes in the insulating phases.

Now, we revisit the condition for adiabatic treatment discussed in Sec.~\ref{secIII}~B. The adiabatic method is justified when the denominator $E_{n,\bm{k}}(\tau)-E_{m,\bm{k}}(\tau)$ is larger than the phonon frequency $\omega$. This holds for sufficiently low frequency $\omega$ in most cases, except when band degeneracy is at the Fermi energy. Thus in our study we can safely use the adiabatic approximation.

\subsection{Why the chiral-phonon-induced current vanishes in insulators?}
As the onsite energy increases, a gap opens between the two bands to become an insulator. In the gapped case, the time-averaged expectation value of the $z$ component of the geometric current Eq.~(\ref{Jinst}) is written as 
\begin{align}
&\braket{j_z^{\rm{geom}}} \nonumber \\
&=\frac{e\omega}{2\pi V}\int_0^{2\pi}d\tau\sum_{\bm{k}}\sum_{n}^{\rm{occ}}\sum_{m(\neq n)}\Bigl\{\frac{\hat{v}_{z,nm}(\bm{k},\tau)A_{mn}(\bm{k},\tau)}{E_{n,\bm{k}}(\tau)-E_{m,\bm{k}}(\tau)}+\rm{C.c}\Bigr\} \nonumber \\
&=\frac{e\omega}{2\pi V}\int_0^{2\pi}d\tau\sum_{\bm{k}}\sum_{n}^{\rm{occ}}2\mathrm{Im}\bigg\langle\frac{\partial\psi_{n,\bm{k}}}{\partial\tau}\bigg |\frac{\partial\psi_{n,\bm{k}}}{\partial k_z}\bigg\rangle \nonumber \\
&=\frac{e\omega}{2\pi}\int\frac{dk_xdk_y}{(2\pi)^2}C(k_x,k_y),
\end{align}
where $C(k_x,k_y)$ is defined as
\begin{equation}
\label{Chern}
C(k_x,k_y)=\sum_{n}^{\rm{occ}}\int_0^{2\pi}d\tau\int_{-\pi}^{\pi}\frac{dk_z}{2\pi} 2\mathrm{Im}\bigg\langle\frac{\partial\psi_{n,\bm{k}}}{\partial\tau}\bigg |\frac{\partial\psi_{n,\bm{k}}}{\partial k_z}\bigg\rangle.
\end{equation}
It is the Chern number for the occupied bands in the parameter $(k_z,\tau)$ space and it is quantized to be an integer for gapped systems. Therefore in gapped systems, $C(k_x,k_y)$ is an integer independent of $k_x$ and $k_y$, and we get
\begin{equation}
\braket{j_z^{\rm{geom}}}=\frac{e\omega}{2\pi S_{xy}}C,
\end{equation}
where $C\equiv C(k_x,k_y)$ is the Chern number, and $S_{xy}$ is the two-dimensional unit cell size within the $xy$ plane. Thus, the time average of the induced current along the $z$ axis is quantized in insulators. The total charge $Q$ carried in one cycle per two-dimensional unit cell within the $xy$ plane is
\begin{equation}
Q=TS_{xy}\braket{j_z^{\rm{geom}}}=Ce,
\end{equation}
which is an integer multiple of the electron charge. Therefore, if the Chern number for the occupied bands is nonzero, it realizes the Thouless pumping~\cite{Thouless} with quantized pumped charge per cycle. 

In our model, however, it vanishes in the insulating phases as shown in Fig.~\ref{cur_on}(a-1), and it indicates that the Chern number $C=0$. To see this, we note that the Chern number can be reinterpreted as a total monopole charge of Weyl nodes in a parameter space in the following way. The Hamiltonian Eq.~(\ref{BH-L}) is written as
\begin{equation}
\mathcal{H}^{L}(\bm{k},t_c,t_s)=\mathcal{H}_0^L(\bm{k})+t_c\mathcal{H}_c(\bm{k})+t_s\mathcal{H}_s(\bm{k}),
\end{equation}
with $t_c=\delta t\cos\omega t$, $t_s=\delta t\sin\omega t$. Then, the Chern number $C$ is the total Berry flux penetrating across the cylinder $t_c^2+t_s^2=\delta t^2$ within the $t_c$-$t_s$-$k_z$ space when $k_x$ and $k_y$ are fixed. It is equal to the total monopole charge of Weyl nodes inside the cylinder. In the present case, the system is gapped everywhere within the cylinder and no Weyl node exists, resulting in $C=0$.

\subsection{Toy model for topological charge pumping}
Apart from the setup of electronic system modulated by chiral phonons, we can propose a realization of the charge pumping with nonzero $C$ by a toy model.
Here we demonstrate how to realize topological charge pumping with the non-trivial Chern number, which is called Thouless pumping~\cite{Thouless}. 
To characterize topological properties of the model Eq.~(\ref{BH-L}), we define $t_c=\delta t_0\cos\omega t$ and $t_s=\delta t_0\sin\omega t$ as time-dependent parameters, and its trajectory is a circle in the $t_c$-$t_s$ plane with a radius $\delta t_0$. 
Then the Hamiltonian is defined in the five-dimensional parameter space $(k_x,k_y,k_z,t_c,t_s)$, and the Chern number $C$ in the $(k_z,\tau)$ plane is the Chern number (i.e.~the total Berry flux) on the cylinder $t_c^2+t_s^2=\delta t_0^2$ within the $t_c$-$t_s$-$k_z$ space. 

As we discussed before, in insulators, the pumped charge per cycle per unit cell within the $xy$ plane is equal to $Ce$, which is an integer multiple of the electron charge. In electronic systems with 
hoppings modulated by atomic motions as studied in the present paper, the Chern number $C$ is zero, and the pumped charge is zero in most cases, and so far we did not find 
any electronic models with atomic motions to realize nontrivial charge pumping. Meanwhile, apart from electronic models with atomic motions, we can find
a three-dimensional model with temporal modulation to realize nontrivial topological charge pumping. 
In this appendix, we propose a toy model with nontrivial charge pumping, i.e.~Thouless pumping, whose Hamiltonian is written as 
\begin{align}
&\mathcal{H}(k_x,k_y,k_z,t_c,t_s)  \nonumber \\
=& \bigg[(1+\cos k_z)t_c-(2-\cos k_x)\bigg]\sigma_x \nonumber + \sin k_z\sigma_y \nonumber \\
&+ \bigg[(1+\cos k_z)t_s-(2-\cos k_y)\bigg]\sigma_z, 
\label{eq:Chern-H}
\end{align}
where $\sigma_i(i=x,y,z)$ are Pauli matrices. As we explain later, it has a nontrivial Chern number $C=1$ on the cylinder 
$t_c^2+t_s^2=\delta t^2_0$ in the $(t_c,t_s,k_z)$ space, when $\delta t_0$ is sufficiently large.
Before discussing the topological properties of the model, we briefly explain how to construct this model. 

\subsubsection{Construction of the model}
\begin{figure}[htb]
%\begin{figure*}[htb]
\begin{center}
\includegraphics[clip,width=8.5cm]{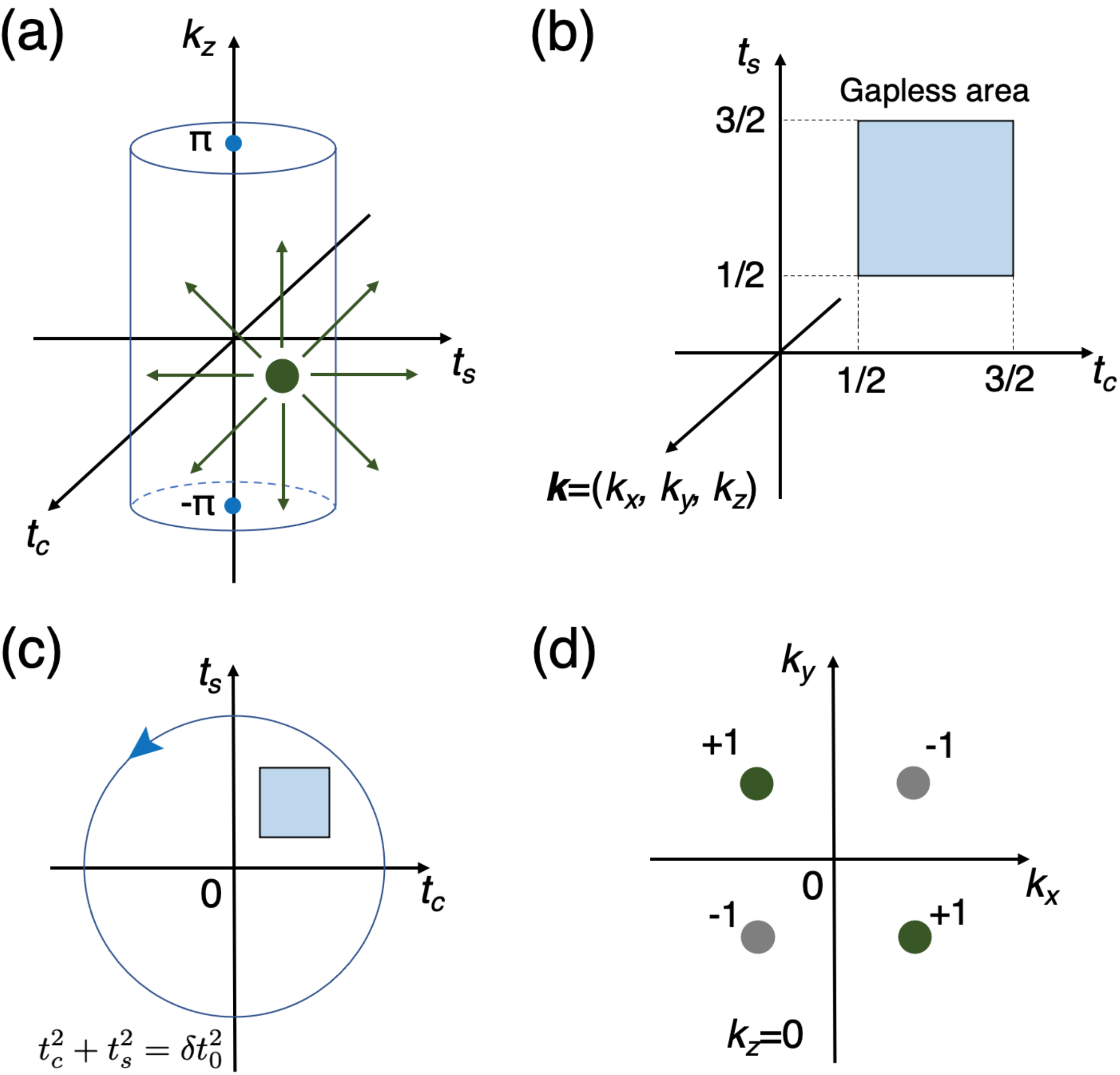}
\end{center}
\caption{Toy model for topological charge pumping. (a) Monopole inside the cylinder at the $k_z=0$ plane in the three-dimensional parameter space $(t_c,t_s,k_z)$. (b) Gapless area in the five-dimensional parameter space $(k_x,k_y,k_z,t_c,t_s)$. (c) Trajectory formed by $t_c$ and $t_s$ encircles the gapless area to realize the charge pumping. (d) Distribution of Weyl points on the $k_z=0$ plane and their monopole charges with $t_c=t_s=1$.}
\label{appendix}
\end{figure}

The Chern number $C$ for the cylinder is equal to the number of monopole charges for Weyl nodes inside the cylinder. Therefore, 
to make the Chern number on the cylinder to be nontrivial, the simplest idea is to have a Weyl node in the 
$(k_z,t_c,t_s)$ space (see Fig.~{\ref{appendix}}(a))
\begin{align}
&\mathcal{H}(k_z,t_c,t_s)  = t_c\sigma_x \nonumber + k_z\sigma_y+ t_s\sigma_z, 
\label{eq:Hkztcts}
\end{align}
which preserves the time-reversal symmetry, $\mathcal{H}^*(k_z,t_c,t_s)=\mathcal{H}(-k_z,t_c,t_s)$.
Since we are considering three-dimensional systems, we need to include $k_x$- and $k_y$-dependence, 
and we can implement it as
\begin{align}
&\mathcal{H}(k_z,t_c,t_s) = (t_c-k_x^2) \sigma_x \nonumber +k_z\sigma_y+ (t_s-k_y^2)\sigma_z, 
\end{align}
to preserve time-reversal symmetry, which leads to be $\mathcal{H}^*(k_x,k_y,k_z,t_c,t_s)=\mathcal{H}(-k_x,-k_y,-k_z,t_c,t_s)$. 
Next, we need to put the system on a lattice, in order to define the Chern number. A 
straighforward way is to make the Hamiltonian as
\begin{align}
&\mathcal{H}(k_z,t_c,t_s)  \nonumber \\
=& \bigg[t_c-(1-\cos k_x)\bigg]\sigma_x \nonumber + \sin k_z\sigma_y+ \bigg[t_s-(1-\cos k_y)\bigg]\sigma_z. 
\end{align}
However, a new problem arises here. Because the gap closes when $\sin k_z=0$, i.e.~$k_z=0,\pi$, there are
two Weyl points with opposite monopole charges, one at $k_z=0$ and the other at $k_z=\pi$, 
within the $k_z$-$t_c$-$t_s$ space. Their contributions to the 
Chern number sum up to zero. To make the Chern number nontrivial, we bring the Weyl point at
$k_z=\pi$ outside of the cylinder, $(t_c,t_s)=(\infty,\infty)$, by introducing a $k_z$-dependence into the coefficients of $t_c$ and 
$t_s$. This leads to the toy model in Eq.~(\ref{eq:Chern-H}).

%---------------------------------------------------
%	Fig. 7: 
%---------------------------------------------------
%\end{figure*}
%---------------------------------------------------

\subsubsection{Topological properties of the model}
Let us discuss topological properties of the model (\ref{eq:Chern-H}). For any fixed value of $k_x$ and $k_y$, the gap closes at $k_z=0,t_c=1-\frac{1}{2}\cos k_x,t_s=1-\frac{1}{2}\cos k_y$, Therefore, there are four Weyl points whose coordinate are $\bm{k}_0^{\alpha\beta}=(\alpha\arccos2(1-t_c),\beta\arccos2(1-t_s),0)$ ($\alpha,\beta=\pm$) with a monopole charge $-\alpha\beta(=1,-1)$,
whose distribution is symmetric due to the broken inversion symmetry and preserved time-reversal symmetry~\cite{NJP}. For example, Fig.~{\ref{appendix}}(d) shows the distribution of the Weyl points on the $k_z=0$ plane and their monopole charges with $t_c=t_s=1$. These Weyl points exist within the parameter range $\frac{1}{2}\leq t_c\leq\frac{3}{2}$ and $\frac{1}{2}\leq t_s\leq\frac{3}{2}$ as shown in Fig.~{\ref{appendix}}(b).  
In the temporal change of the parameters $t_c$ and $t_s$ with $t_s=\delta t_0\sin\tau$ and $t_c=\delta t_0\cos\tau$, their trajectory forms a cycle in the $t_c$-$t_s$ plane as shown in Fig.~{\ref{appendix}}(c). Once the trajectory encircles the gapless area (Fig.~{\ref{appendix}}(b)) for a given radius $\delta t_0$, for any value of $k_x$ and $k_y$, there is only one monopole inside the cylinder as shown in Fig.~{\ref{appendix}}(a). Therefore, the Chern number defined on the cylinder in Eq.~({\ref{Chern}}) is equal to one. 

From these considerations, it is difficult to achieve Thouless charge pumping in the present setup of electronic systems with chiral phonons. Because its realization requires nonzero $C\equiv C(k_x,k_y)$ for every value of $(k_x,k_y)$, the gap should be closed somewhere inside the cylinder for every value of $(k_x,k_y)$. It means that the energy modulation by the change of parameters $(t_c,t_s)$ should be larger than the bandwidth. Meanwhile, the phonon energy is usually much smaller than the bandwidth, and therefore, the Thouless charge pumping cannot be realized by chiral phonons.

\section{CONCLUSION}
\label{secV}
In summary, we consider a microscopic local rotation of atoms in a helical structure with the honeycomb-lattice layers. By treating the rotational motion as a time-dependent perturbation, we calculate the time-dependent current by using the Berry phase method under the adiabatic approximation. The results show that the time-averaged current is separated into two parts: the instantaneous current and the geometric current. The instantaneous current is trivially given by the diagonal instantaneous matrix element of the velocity which can be viewed as a snapshot process, and this term vanishes under the time-reversal symmetry. The geometric current is expressed as a product of the instantaneous matrix element for the velocity and the Berry connection, and is proportional to the phonon frequency $\omega$. We calculate the geometric current induced by chiral phonons in the metallic and insulating phases. The time average of the current becomes finite along the helical axis in the metallic phases. However it vanishes when the onsite energy is large to become an insulator. On the other hand, in the hexagonal plane, the current changes with time, but the time average of the current vanishes due to the space-time threefold rotation symmetry. 

In this study, our model is intended to be a minimal model to demonstrate the chiral-phonon-induced current, so we retain only a few terms necessary for our discussion. We have shown that even in a simple toy model with helical structure, the chiral phonons induce a current. Actually, in real materials, the crystal structure and the electronic bands are much more complex than the toy model used in this paper. Our theoretical prediction is generally applied to real materials, and chiral-phonon-induced current is expected in real materials with chiral crystal structure.

In the present paper, we studied the effect of chiral phonons at the $\bm{\Gamma}$ point. In thermal equilibrium, however, the $\bm{\Gamma}$-point phonons are doubly degenerate between chiral phonons with opposite handedness, which makes the net current zero. In order to get a net nonzero current, one would need an asymmetric population of chiral phonons between the right-handed and left-handed circular polarizations. This could for example be achieved through phonon pumping with terahertz pulses~\cite{GaCP1}, or through some other mechanisms. On the other hand, a natural question arises as to an effect of chiral phonons at general non-$\bm{\Gamma}$ high-symmetry points. In fact, our results in the present paper have some implications on the effect of chiral phonons at general non-$\bm{\Gamma}$ points. Chiral phonons at general non-$\bm{\Gamma}$ high-symmetry points can always be brought to the $\bm{\Gamma}$ point by appropriately enlarging the unit cell. Thereby, the above results similarly hold in such cases. Details of the properties of the chiral-phonon-induced current due to the non-$\bm{\Gamma}$ chiral phonons are left as future works.

\begin{acknowledgments}

This work was partly supported by JSPS KAKENHI Grants No. JP18H03678 and JP22H00108.
\end{acknowledgments}

\begin{appendix}

\section{Derivation of the helical Bloch Hamiltonian}
In this Appendix, we give more details on the derivation of the helical tight-binding Hamiltonian Eq.~(\ref{HR}), and the connection between Eq.~(\ref{HR}) and its the Bloch Hamiltonian Eq.~(\ref{BlochHam}).

First of all, we note that although the positions of the lattice sites look like a non-helical one as shown in Fig~\ref{Struct}, in general, the positions of the lattice sites in the tight-binding models are not necessarily the same as the the positions of all the atoms. We assume the crystal structure, i.e. the atomic positions to be chiral, whereas the lattice sites are located in a non-chiral way, which leads to two patterns, the left-handed and right-handed helices (see Figs.~\ref{Struct}(b) and~\ref{Struct}(c)). This helical hopping along the $z$ direction in Eq.~(\ref{HR}) is specifically expressed as 
\begin{align}
\hat{H}_{0,\rm{2nd}}^{L,R}=\sum_{\bm{r}}\sum_{i=1}^{3}&t_A\hat{c}_{A,\bm{r}}^{\dagger}\hat{c}_{A,\bm{r}\pm\bm{b}_i+\bm{a}_3}+ \nonumber \\
&t_B\hat{c}_{B,\bm{r}}^{\dagger}\hat{c}_{B,\bm{r}\mp\bm{b}_i+\bm{a}_3}+\rm{h.c}, \nonumber
\end{align}
where vectors $\bm{b}_i$ are defined in Sec.~\ref{secII}~A. 
 
Applying the Fourier transformation
\begin{align}
&\hat{c}_{\mu,\bm{r}}=\frac{1}{\sqrt{N}}\sum_{\bm{k}}\hat{c}_{\mu,\bm{k}}e^{i\bm{k}\cdot\bm{r}}, \\
&\hat{c}^{\dagger}_{\mu,\bm{r}}=\frac{1}{\sqrt{N}}\sum_{\bm{k}}\hat{c}^{\dagger}_{\mu,\bm{k}}e^{-i\bm{k}\cdot\bm{r}}, 
\end{align}
to Eq.~(\ref{HR}) for the two sublattices $\mu=\rm{A,B}$, the matrix representation of the Bloch Hamiltonian Eq.~(\ref{BlochHam}) for the left-handed helix is 
\begin{widetext}
\begin{eqnarray}
\mathcal{H}_0^{L}(\bm{k})=
\begin{pmatrix}
\lambda_{\nu}+2t_{A}\sum_{i=1}^{3}\cos{[\bm{k}\cdot (\bm{b}_i+\bm{a}_3)]} & t_1(1+e^{-i\bm{k}\cdot\bm{a}_1}+e^{-i\bm{k}\cdot\bm{a}_2}) \\
t_1(1+e^{i\bm{k}\cdot\bm{a}_1}+e^{i\bm{k}\cdot\bm{a}_2}) & -\lambda_{\nu}+2t_{B}\sum_{i=1}^{3}\cos{[\bm{k}\cdot (-\bm{b}_i+\bm{a}_3)]} \\
\end{pmatrix},
\end{eqnarray}
\end{widetext}
and for the right-handed helix is 
\begin{widetext}
\begin{eqnarray}
\mathcal{H}_0^{R}(\bm{k})=
\begin{pmatrix}
\lambda_{\nu}+2t_{A}\sum_{i=1}^{3}\cos{[\bm{k}\cdot (-\bm{b}_i+\bm{a}_3)]} & t_1(1+e^{-i\bm{k}\cdot\bm{a}_1}+e^{-i\bm{k}\cdot\bm{a}_2}) \\
t_1(1+e^{i\bm{k}\cdot\bm{a}_1}+e^{i\bm{k}\cdot\bm{a}_2}) & -\lambda_{\nu}+2t_{B}\sum_{i=1}^{3}\cos{[\bm{k}\cdot (\bm{b}_i+\bm{a}_3)]} \\
\end{pmatrix}.
\end{eqnarray}
\end{widetext}
Some parameters shown here have already been defined in the main text. Here we can see that the helical hopping term along the $z$ direction appears between the same sublattices, which corresponds to the second term in the tight-binding Hamiltonian Eq.~(\ref{HR}).

\end{appendix}

\end{document}